\documentclass{article}

\usepackage{multirow}
\usepackage{graphicx}
\usepackage{wrapfig, lipsum}

\usepackage[preprint]{neurips_2023}

\usepackage{natbib}
\usepackage[utf8]{inputenc} 
\usepackage[T1]{fontenc}    
\usepackage{hyperref}       
\usepackage{url}            
\usepackage{booktabs}       
\usepackage{amsfonts}       
\usepackage{nicefrac}       
\usepackage{microtype}      
\usepackage{graphicx}
\usepackage{soul}
\usepackage[dvipsnames]{xcolor}   
 \makeatletter 
 \newcommand\citecolor[1]{\@namedef{keycolor#1}{\color{red}}}
 \makeatother
 \citecolor{Schmidt2011}

\title{ECGBERT: Understanding Hidden Language of ECGs with Self-Supervised Representation Learning}

\author{%
Seokmin Choi$^{1, 2,\dag}$\quad Sajad Mousavi$^{1,\dag}$\thanks{Corresponding author.  \dag Equal contribution.} \quad Phillip Si$^{1, 3,\dag}$ \\
\quad \textbf{Haben G. Yhdego}$^{1}$ \quad \textbf{Fatemeh Khadem}$^{1}$ \quad \textbf{Fatemeh Afghah}$^{4}$ \\
$^{1}$\textit{CardioPhi LLC}, CA, USA\\
$^{2}$University at Buffalo, SUNY, NY, USA\\
$^{3}$Carnegie Mellon University, PA, USA\\
$^{4}$Clemson University, SC, USA\\
\texttt{\{seokmin.choi,sajad.mousavi,phillip.si\}@cardiophi.com}\\
\texttt{\{haben.yhdego,fatemeh.khadem\}@cardiophi.com}\\
\texttt{fatemeh.afghah@clemson.edu}\\
}

\begin{document}

\maketitle

\begin{abstract}
In the medical field, current ECG signal analysis approaches rely on supervised deep neural networks trained for specific tasks that require substantial amounts of labeled data. However, our paper introduces ECGBERT, a self-supervised representation learning approach that unlocks the underlying language of ECGs. By unsupervised pre-training of the model, we mitigate challenges posed by the lack of well-labeled and curated medical data. ECGBERT, inspired by advances in the area of natural language processing and large language models, can be fine-tuned with minimal additional layers for various ECG-based problems. Through four tasks, including Atrial Fibrillation arrhythmia detection, heartbeat classification, sleep apnea detection, and user authentication, we demonstrate ECGBERT's potential to achieve state-of-the-art results on a wide variety of tasks.
\end{abstract}

\section{Introduction}
The Centers for Disease Control (CDC) reported that heart disease is the leading cause of death in the United States \citep{cdc_report22}. Specifically, one person dies every 34 seconds from cardiovascular disease and about 697,000 people died from heart disease in 2020, which is one in every five deaths. The electrocardiogram (ECG) is the most essential bio-signal used by cardiologists and physicians to keep track of heart activity and detect different heart-related diseases and is used by cardiologists and physicians. One of the most critical limitations of ECG signals is that it requires manual analysis and annotation. Furthermore, the interpretation of the ECG signals varies from physician to physician as different heart diseases are associated with complex patterns within the ECG which can be hard to detect. The resulting inconsistencies may affect diagnostic accuracy or the trust between the patient and the physician. Therefore, to mitigate the aforementioned limitations in regard to manual ECG interpretation, several studies have proposed alternative ECG analysis techniques to achieve higher accuracy in real-time. Among these, deep learning-based approaches have recently gained traction in this domain \citep{pyakillya2017deep,mousavi2020han}. Compared with machine learning-based approaches where features need to be extracted manually, deep learning-based approaches automatically extract relevant features \citep{rajpurkar2017cardiologist,isin2017cardiac}, allowing for improved performance given enough data and a sufficiently expressive model. As a result, deep learning techniques have been widely applied to the medical domain in recent years to solve different medical-related problems such as Atrial Fibrillation (AFIB) arrhythmia detection or heartbeat classification \citet{mousavi2019inter, andersen2019deep}. 

However, even with a large amount of data and sufficient computation, deep learning models are typically designed for specific tasks, limiting their applicability to one task at a time. Achieving optimal performance with deep learning models necessitates a substantial amount of data, which is a critical challenge in the medical field due to privacy constraints and data availability. While the curated and well-labeled ECG data is limited, there is a wealth of unlabeled ECG data that remains untapped. In addition, deep learning models, which are constructed by stacking multiple layers, require a large number of learnable parameters and extensive data as the model becomes deeper. Even when these complications have been resolved, the resulting models are tailored to their specific task and lack versatility for broader applications. Therefore, despite the fact that previous ECG-related models have shown promising results, the methods have limited applicability in the real world. To overcome these limitations of deep learning-based methods, it is necessary to design a more versatile and universal model that also resolves the data label or annotation issue. 

A large language model (LLM) is a machine learning model within the field of natural language processing (NLP) that is capable of processing and emulating human language. LLMs are trained on a vast amount of text data and exhibit remarkable capabilities in diverse NLP tasks, including summarization \citep{cai2021chestxraybert}, text generation \citep{kumar2021controlled}, sentiment analysis \citep{liu2012emoticon}, or question-answering \citep{khot2020qasc}. Among the most famous LLMs currently are BERT \citep{devlin2018bert}, GPT and its variants  \citep{radford2018improving}, \citep{brown2020language}, and transformer-based architecture models \citep{vaswani2017attention}. Given the outstanding performance demonstrated by LLMs in NLP, researchers have explored the possibility of extrapolating LLM models to other domains. For example, \citet{khan2022transformers} introduced vision transformers which divide an image into smaller patches, the equivalent of words in NLP. Transformers have also been applied to a variety of robotics problems \citep{brohan2022rt}, and LLMs were used in the field of law as well \citep{choi2023chatgpt,nay2023large}.



Recently, \citet{mousavi2021ecg} introduced a novel method called ECG language processing (ELP), which applies NLP-style models to analyze ECG signals. While traditional models commonly employed in ECG analysis are limited by their reliance on labeled data, the medical field possesses an extensive repository of unrefined and unlabeled ECG records. This parallelism with the corpus of textual data highlights the vast number of unrefined and unlabeled ECG records within the medical field, which remains untapped by conventional models designed for ECG analysis.

\begin{figure}[h]
    \centering
    \includegraphics[scale=0.75]{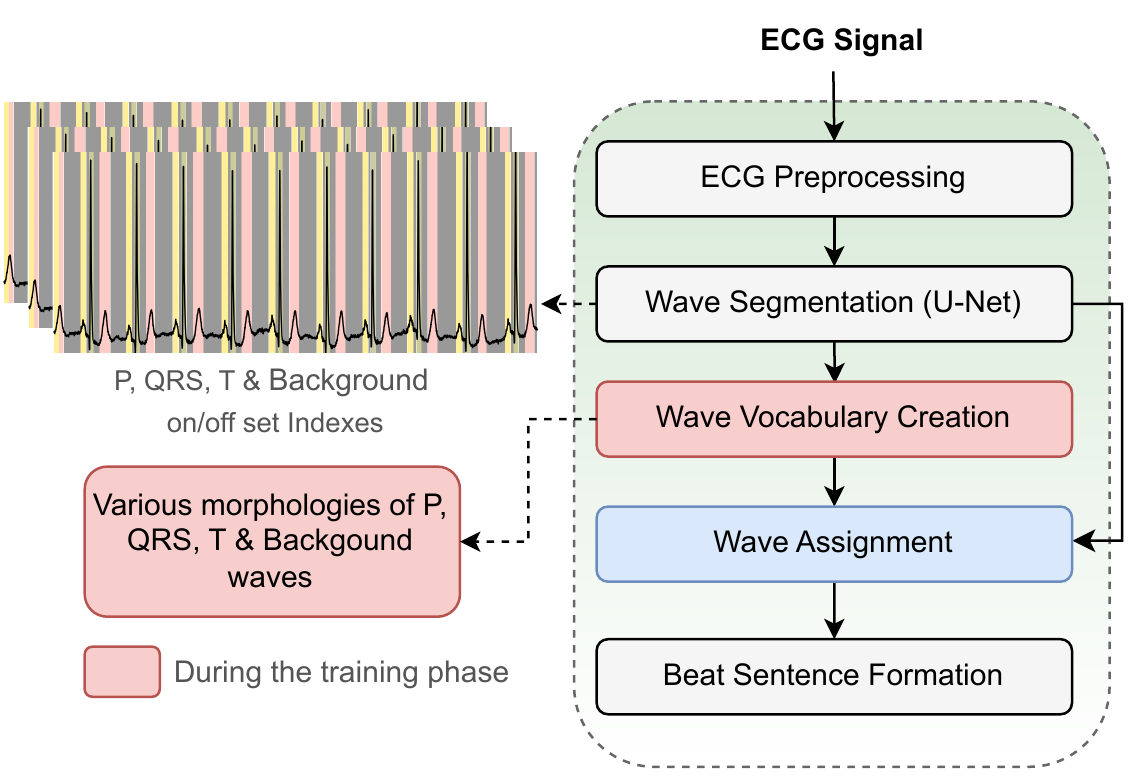}
    \caption{Holistic view of ECG language processing (ELP)}
    \label{fig:flowchart}
\end{figure}

\begin{figure*}[!t]
    \centering
    \includegraphics[scale=0.62]{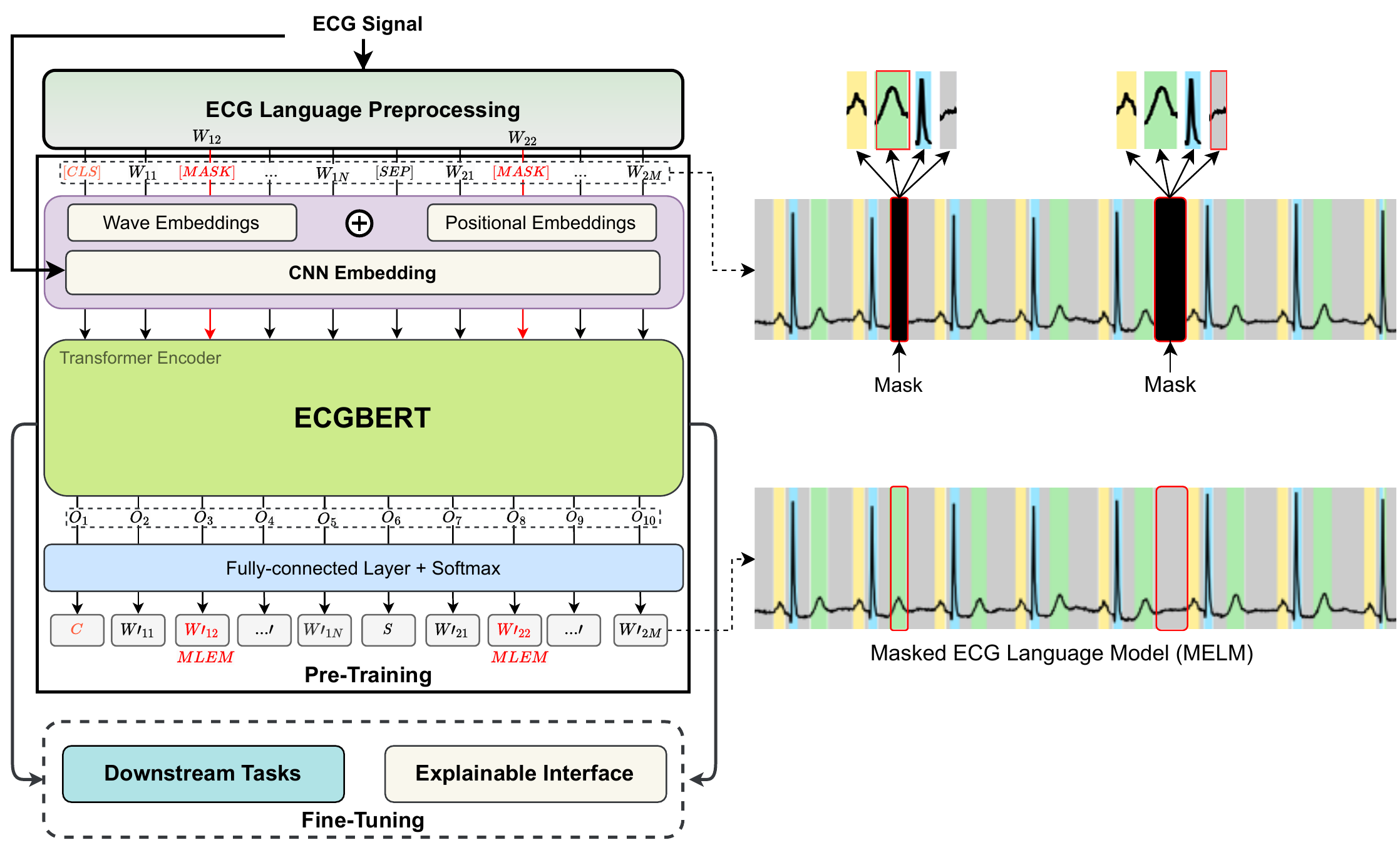}
    \caption{Illustration of the proposed ECGBERT architecture}
    \label{fig:ECGBERT}
\end{figure*}

To address this challenge, we introduce ECGBERT, a novel LLM model framework inspired by BERT \citep{devlin2018bert}. ECGBERT capitalizes on large amounts of unlabeled ECG data during the pre-training stage to learn meaningful representations such that downstream tasks can be adapted efficiently with a minimal amount of labeled data. To achieve this, we integrate the framework for BERT (Bidirectional Encoder Representations from Transformers), an LLM used within the NLP domain, with the upgraded version of ELP paradigm, as shown in Figure \ref{fig:flowchart}. \ref{fig:ECGBERT} illustrates a detailed overview of the model architecture.

The proposed framework allows us to create a versatile and potent tool applicable to various medical tasks including, but not limited to, heartbeat classification, cardiac arrhythmias detection, sleep Apnea detection, or even user authentication, all of which rely on ECGs as the input. Because we pre-train ECGBERT on a large amount of ECG data in an unsupervised manner, this enables the model to learn and represent the nuances, complexities, and latent patterns of ECG signals without the need for human supervision or annotation. This unique capability of ECGBERT facilitates a more efficient and effective analysis and interpretation of ECG signals.

Our proposed framework's contributions are multifold: Firstly, it analyzes and details the method of mapping between textual tokens within the NLP domain and \textbf{continuous} ECG signals. Secondly, it permits learning more general representations of ECG signals while still capturing subtler fine-grained pattern differences after the transformation into tokens. Thirdly, it offers embedded explainability and interpretability as it is treating ECG signals like text within the NLP domain, shedding light on the model's decision-making process. To our best understanding, this is the first paper that truly re-interprets ECG signals from an unsupervised LLM perspective.

Our paper is structured in the following manner:
\begin{itemize}
    \vspace{-3pt}
    \item We interpret the time series ECG signals as integer-encoded ECG tokens by creating a wave vocabulary and wave assignment.
    \vspace{-4pt}
    \item We utilize a CNN encoder to create shift-invariant embeddings to capture the finer-grained characteristics of the ECG signals in combination with general ELP tokens.
    \vspace{-4pt}
    \item We introduce ECGBERT, a novel deep learning model that combines state-of-the-art ideas from both ECG and NLP domains, that can be applied to various downstream tasks after it is pre-trained in an unsupervised manner (i.e., unlabeled data).
    \item We show that ECGBERT performs competitively across four distinct downstream tasks, demonstrating its versatility and effectiveness.
\end{itemize}

\section{Related Work} 

BERT, the primary inspiration for our work, is based on a transformer architecture that consists of attention blocks where each output is connected to its input and determines the importance of their relationship \citep{vaswani2017attention}. One of the most essential characteristics of BERT is the departure of heavy reliance on well-refined labels during pre-training, which is especially suitable for medical data due to the myriad privacy restrictions. Moreover, BERT was proposed to resolve one of the critical limitations of the previous language models: uni-directionality, which can only leverage the previous tokens in the attention layers \citep{radford2018improving}, resulting in suboptimal performance on downstream tasks which requires utilizing context from both directions such as question answering. 


\subsection{ECG with Deep Learning} 
As deep learning methods approach human-level accuracy while far surpassing their human counterparts in predictive speed, researchers have begun to apply deep learning techniques to resolve various tasks using ECG data, such as heartbeat classification \citep{mathews2018novel}, heart arrhythmia detection \citep{singh2018classification}, interpretable AFIB classification \citep{nankani2022atrial}, cardiovascular disease diagnosis \citep{qiu2023transfer}, and sleep apnea \citep{feng2020sleep}, with each of them leveraging different architecture styles ranging from dense networks to RNNs to HMMs.



However, all of these studies are task-specific and do not seek to learn a general representation of the ECG language. In this study, we seek to create a framework that can extract good representations for ECG signals which can also utilize context effectively.


\subsection{Large Language Model in the Medical Domain}
Due to the LLMs' outstanding performance in NLP, researchers have tried deploying LLMs to the medical domain due to a range of benefits. Firstly, to improve the accuracy and reliability due to the ability of LLMs to learn the relationships between different medical conditions and symptoms. Secondly, helping interpret medical tests such as laboratory results with explainable attention weights. Finally, LLMs can possibly be applied to identify clinical trials which may be relevant to the patient's health condition. GatorTron \citep{yang2022large} develops a large clinical language model using more than 90 billion words of text, and performs five clinical NLP tasks such as medical relation extraction or semantic textual similarity. BEHRT \citep{li2020behrt} introduces a deep sequence transduction model for electronic health records (EHR) which is capable of predicting the likelihood of 301 conditions in future visits. Moreover, MED-BERT \citep{rasmy2021med} adopts the BERT framework to pre-train with structured EHR data and implements two disease prediction tasks from two clinical databases. BioBERT \citep{lee2020biobert} leverages a biomedical domain corpora to pre-train the model and conducts different biomedical text mining tasks (e.g, question answering or next entity recognition) to validate the model's effectiveness. Although the aforementioned studies show promising results, all of them leveraged either the clinical notes or EHR data, which is still textual data. However, ECGBERT employs time series ECG signals, which are transformed into words and vectors to be fed into the proposed model. 
\subsection{Transfer Learning from Supervised ECG Data}
Transfer learning has become an increasingly popular area of research in various domains of machine learning like NLP, and computer vision. It has also shown promising results for ECG analysis using supervised data such as ECG arrhythmia classification \citep{salem2018ecg,strodthoff2020deep,weimann2021transfer} and transferable representation \citep{kachuee2018ecg}. In addition, \citet{li2021transfer} proposed a photoplethysmography (PPG) and actigraphy-based sleep stage classification by applying a model trained on an ECG dataset, and \citet{zhang2021heartbeats} introduced heartbeat classification by adopting a transfer learning technique with a Residual Neural Network (ResNet) that was pre-trained on ImageNet to analyze ECG-signals which were transformed into 2-D time-frequency diagrams. It can be seen that ECG-related tasks are highly amenable to transfer-based learning. We adopt this in an unsupervised manner by pre-training ECGBERT and fine-tuning the model on different downstream tasks.
 
\section{Methodology}
ECGBERT pipeline is composed of two macroscopic tasks: ECG language processing and model training. In particular, the former is mainly composed of five modules: ECG preprocessing, wave segmentation, wave vocabulary creation, wave assignment, and beat sentence formation.

\subsection{ECG Language Processing} \label{sec:elp}
\subsubsection{ECG Preprocessing}
Since raw ECG signals are contaminated during the recording by various factors such as motion artifacts or powerline interference, it is essential to polish the raw data to be cleaner by applying different signal processing techniques. Specifically, in the ECG signals, power-line interference and baseline wander are the two major factors that induce ECG signal corruption \citep{mian2020baseline}. First, to eliminate the powerline interference from the ECG signals, we apply a second-order Butterworth band-stop filter with cut-off frequencies at 50 Hz and 60 Hz. After mitigating the powerline interference issue, we remove the baseline wander by leveraging discrete wavelet transforms (DWT). The objective of the DWT is to decompose the signal into different resolutions using high-pass and low-pass components. In this study, we decompose the signal into one level using the Daubechies 4 wavelet family. We shift and calculate the energy of the detail coefficients, and the baseline is then reconstructed from this level using low-pass signals, as baseline wander is a low-frequency artifact.

\subsubsection{Wave Segmentation} \label{sec:wave_seg}
Segmenting  heartbeats of an ECG signal can drastically affect the diagnosis accuracy. Specifically, in this study, we divide ECG signals into P waves, T waves, QRS complexes, and background waves. Even if parts of an ECG signal belong to the same wave, due to different external factors such as motion artifacts and noises, it can be challenging to extract the correct segmented waves. To segment the ECG signals into a set of different waves, we first adopt the Hamilton algorithm \citep{hamilton2002open} to clean the ECG signals to improve the quality of R-peak detection. We then apply DWT to separate the different waves of the cardiac cycle: 1) the P wave, which is a depolarization wave that spreads throughout the atria, 2) the QRS complex, which has a larger amplitude than other waves and shows a rapid depolarization of both verticles, 3) the T wave, which reflects the ventricular repolarization of the ventricles, and 4) background waves, which don't belong to any of three waves. Based on the time-frequency analysis of the ECG signals, wave segmentation will return the onset and offset indices of each wave.

\subsubsection{Wave Vocabulary Creation} \label{sec:vocab_creation}
ECG Patterns within a given wave group may exhibit variations, even though they belong to the same group. Within the same wave group, different wave types represent different phases of heart activities: electrical activity towards or away from a lead that causes an upward or downward deflection respectively. In particular, heart-related diseases can be identified by interpreting the unique morphology of the waves. 

According to a previous study \citep{ecgWaves}, P waves can be represented by five different morphologies, QRS waves can be represented by twelve morphologies and T waves can be categorized as seven morphologies. As the proposed system is trained to interpret and analyze different wave types to learn the representations of the ECG signals, it is critical to categorize the waves into different groups. So, we employ a clustering model to create a comprehensive ECG wave vocabulary. However, it will be time and labor-intensive if clustering the waves requires the morphology labels. Therefore, we adopt an unsupervised clustering algorithm (i.e., Kmeans) that employs Dynamic Time Warping (DTW) \citep{berndt1994using} to complete this step.
DTW was introduced to measure the similarity between two given temporal sequences which may vary in speed. One of the main advantages of applying DTW is that it shows reliable time alignment when there are two similar patterns with different duration. This is especially crucial for the ECG signals considering the similarity in patterns despite the difference in each patient's heartbeat cycle. 
To cluster the ECG waves, we train four different clustering models to categorize the P waves into 12 clusters, QRS waves into 19 clusters, T waves into 14 clusters, and background waves into 25 clusters (70 clusters in total).
Since different wave morphologies carry different chunks of information about the corresponding ECG signal, we reinterpret the waves as an analogous component to words within the NLP domain. Therefore, the clustering algorithm could be viewed as a method that groups words together based on their semantic similarity to create a wave vocabulary.

\subsubsection{Wave Assignment}
Afterward, each segmented wave is fed into a different clustering model depending on the wave type to assign the wave to the proper corresponding cluster. This encodes the ECG signal into a sequence of integer tokens.

\subsubsection{Beat Sentence Formulation}
Based on the encoded beat waves, we then construct a sentence that consists of one or more heartbeats. In particular, to enhance the generalizability during pre-training, a sentence is formulated by either one, two, six, or eight consecutive heartbeats randomly, where a heartbeat is comprised of multiple consecutive waves. When constructing a sentence composed of one or more heartbeats, we need to determine the onset and offset of the heartbeat wave. Normally for the ECG interpretations, first the P-wave is interpreted, followed by the PR wave in between the offset of the P wave and the onset of the next QRS wave. In this study, the PR wave is categorized as one of the background waves. Afterward, the QRS wave is interpreted, and then the T wave. In summary, the order of a beat sentence starts with P, QRS, and T, with background clusters in between the defined wave clusters.



\subsection{Bidirectional Transformer}
\subsubsection{Input Embedding Representations} \label{sec:input_repre}
After constructing an ECG sentence from the clustered waves, these sentences are then fed into the Bidirectional Transformer part of the pipeline, which is shown in Figure \ref{fig:ECGBERT}. For the input representations, we include positional embeddings, token embeddings, and CNN embeddings.

\paragraph{Positional Embeddings}
Positional embeddings are added to assign orders to the non-recurrent multihead attention, creating a temporal context for the tokens. This is especially critical to ECGBERT which tokenizes the heartbeat wave because the ECG record loses some relevant temporal information during the tokenizing step.

\paragraph{Token Embeddings}
Token embeddings convert heartbeat waves into different tokens. In particular, for the tokenizer, we leverage our own predicted clusters from the wave vocabulary creation module to represent a heartbeat sentence in a single set of tokens.  
Furthermore, the \texttt{[SEP]} token is added at the end of the sentence to serve as a marker indicating the end of the sentence.

\paragraph{CNN Embeddings}
Unfortunately, directly tokenizing the ECG signals and applying BERT's schematic suffers from the inability to capture finer-grained details of ECG signals. While token embeddings provide general representations for each token in input text by capturing the meaning of each word in context and positional embeddings reflect the position of each token in the sequence, the continuous time-series data structure of ECG signals requires much more refined input representations to fully capture subtle differences in various ECG patterns. To address this limitation, we introduce a CNN token embedding that serves as a feature extractor of raw ECG signals. Specifically, we adopt a U-Net architecture with two downsampling and upsampling blocks, along with skip connections and batch normalization layers. The output of the CNN feature extractor is then segmented based on the onset and offset indices of each wave and added with token and positional embeddings.

\subsubsection{Model Architecture}

The model portion of ECGBERT adopts a transformer-encoder-style architecture. The main rationale behind applying attention from transformer architectures is to focus on appropriate parts of the ECG sequence and determine the important neighboring components by applying scaled dot-product operation. In addition, we also adopt multi-headed attention layers in this framework.


\subsubsection{Pre-training ECGBERT}
Similar to BERT, we apply  Masked Language Modeling (MLM) unsupervised technique during the pre-training stage. We apply MLM to train a bidirectional representation of the ECG signals which helps the model learn the temporal relationships between nearby heartbeats. As it can be seen from the two right figures in Figure \ref{fig:ECGBERT}, MLM is implemented by masking 15\% of the wave tokens randomly and asking ECGBERT to predict the masked waves. We didn't include Next Sentence Prediction (NSP) task, as a necessity of the NSP is not as crucial as originally thought which was discussed in some of the previous studies\citep{cui2021pre,tinn2023fine}. In particular, the NSP task only requires the model to predict whether two given sentences are consecutive or not, which may not be a good representation of the subtle but complex patterns that exist in ECG signals.

\subsubsection{Pre-training Datasets}
Since ECGBERT adopts an unsupervised learning approach during the pre-training stage where the model learns the general representations of ECG signals, utilizing ECG datasets without labels is one of the main contributions of our proposed system which can reduce the cost and time. For the pre-training purpose, we adopt MIMIC-III waveform \citep{moody2020mimic}, PTB-XL \citep{wagner2020ptb}, Georgia \citep{alday2020classification}, CPSC-2018 \citep{liu2018open} datasets which can be downloaded from the Physionet website \citep{goldberger2000physiobank}. We then pretrain our proposed ECGBERT using two RTX 2080Ti GPUs on a local machine, with around 236 hours of data due to the limited computing power. With more data and computing power, we expect the ECGBERT to learn even better representations than the model we used for our downstream tasks.



\subsubsection{Fine-Tuning ECGBERT}
Following the pre-training phase, the pre-trained ECGBERT model is utilized in diverse downstream applications, extending beyond medical tasks such as AFIB or heart arrhythmia detection. The potential of deploying ECGBERT for alternative tasks like sleep apnea detection or even user authentication has also been investigated. This low-cost fine-tuning process involves adding one or two additional dense layers on top of the pre-trained ECGBERT model, which can be achieved by adjusting the model weights. This process will be further detailed in the subsequent section.

\section{Experiments}
This section presents four different downstream tasks using a pre-trained ECGBERT from the previous section. For evaluation metrics, we adopt accuracy, specificity, sensitivity, and  positive predictive value (PPV). For the datasets for which other papers have different data setups (e.g. different input lengths or inter/intra patient setups), we benchmark against a ResNet model containing three residual blocks with ReLU activations before a final linear layer. Batchnorms are applied after every convolutional layer within each of the blocks. The details for the publicly available datasets used for training and evaluation can be found in Appendix \ref{sec:downstream_data}. In this study, we focus on the inter-patient evaluation scheme, which reflects a more realistic scenario compared to the commonly employed intra-patient paradigm. In an inter-patient experimental design, train and test data are divided at a patient level before splitting into sub-segments. On the other hand, an intra-patient scheme divides the data into small segments first before randomly assigning these small segments to train or test, which allows data from the same patient to be in both train and test sets. As a result, train and test distributions are much more similar to an intra-patient schematic, which is unrealistic in real-world scenarios. By avoiding biases introduced by training and testing on the same patient's samples, our results reflect a more reliable comparison with existing methods \citep{de2004automatic}.

\begin{table}[]
\centering
\caption{AFIB Arrhythmia Detection Performance Comparison with State-of-the-Art Methods \newline (RRI: RR-interval)}
\label{tab:afib_perf}
\resizebox{\textwidth}{!}{%
\begin{tabular}{lcccccc}
\hline
    \multicolumn{1}{c}{\textbf{Model}}                                     & \textbf{Paradigm} & \textbf{Signal Length} & \multicolumn{4}{c}{\textbf{Performance}}                                         \\
\cline{4-7} 
\multicolumn{1}{c}{}  &                   &                        & \textit{\textbf{Accuracy}} & \textit{\textbf{Specificity}} & \textit{\textbf{Sensitivity}} & \textit{\textbf{PPV}}   \\ \hline

\citet{tuboly2021atrial}                                                                     & Intra-patient     & 60s                    & 0.980             & 0.987                & 0.974                & 0.988          \\ \hline

ResNet                                                                     & Inter-patient     & 10s         &     0.884           &       0.951     &     0.846          &    0.969      \\
 \citet{andersen2019deep} &    Inter-patient        &  30 RRIs                 &   \textbf{0.978}           & \textbf{0.989}                & 0.969                &   0.957           \\
\citet{pereira2022inter} & Inter-patient     & 10                     & 0.908             & 0.910                & 0.915                & -              \\

\textbf{ECGBERT}                & Inter-patient     & 10s                    &0.973    & 0.976       & \textbf{0.970}       & \textbf{0.981} \\ \hline
\bottomrule
\end{tabular}%
}
\end{table}

\subsection{AFIB Arrhythmia Detection}

For this downstream task, we explore the performance of the pre-trained ECGBERT model on the detection of AFIB rhythms by leveraging the MIT-BIH Atrial Fibrillation database \citep{moody1983new}.

In this study, we fine-tune the model with two extra dense layers on top of the ECGBERT with a learning rate of 0.001 and a batch size of 64  for 13 epochs. As in Table \ref{tab:afib_perf}, ECGBERT achieves an accuracy of 0.973, a specificity of 0.976, a sensitivity of 0.970, and a PPV of 0.981. Note that we only compared the performance with previous studies that evaluated the performance based on the inter-patient approach (i.e., \citep{andersen2019deep,pereira2022inter}). Even though some studies may outperform ECGBERT (e.g., \citep{tuboly2021atrial}) their approaches seem to be evaluated on intra-patient data, which is unrealistic as similar distributions at train and test time are not guaranteed under real-world circumstances. Despite these noted advantages, however, ECGBERT can still achieve comparable performance to those studies. 

\subsection{Heartbeat Classification}
Next, we fine-tune ECGBERT to perform Heartbeat Classification by leveraging the MIT-BIH Arrhythmia database  \citep{moody2001mitbih} with an inter-patient paradigm. Among different beat types, we divide the beats into five classes: normal (N), unknown (Q), supraventricular ectopic (S), ventricular ectopic (V), and fusion (F) heartbeat groups recommended by the American Association of Medical Instrumentation (AAMI) \citep{ec571998testing}. For finetuning layers, we add two residual blocks on top of BERT to capture a variable number of labels to predict. We finetune the model with a learning rate of 1e-4 for 20 epochs.

From Table \ref{tab:multiclass}, we can see that ECGBERT offers a reasonable performance which still has room for improvement since beat classification tasks need to have appropriate masking and data organization in order to classify a variable number of beats within a 10s ECG segment.

\begin{table}[!t]
\centering
\caption{Confusion Matrix and Per-Class Performance of Heartbeat Classification Implemented by ECGBERT on the MIT-BIH Database.}
\label{tab:multiclass}
\begin{tabular}{cccccclcccc}
\hline
              &            & \multicolumn{4}{c}{\textbf{Predicted}}            &  & \multicolumn{4}{c}{\textbf{Per-class Performance}}         \\ \cline{3-6} \cline{8-11} 
\textbf{}     &            & \textbf{N} & \textbf{S} & \textbf{V} & \textbf{Q} &  & \textit{\textbf{Accuracy}} & \textit{\textbf{Specificity}} & \textit{\textbf{Sensitivity}} & \textit{\textbf{PPV}} \\ \hline
\textbf{True} & \textbf{N} & 38538         & 1483          & 1941          & 1119          &  & 0.86         & 0.45          & 0.89         & 0.94         \\
              & \textbf{S} & 187          & 26          & 39          & 7          &  & 0.95         & 0.99          & 0.10         & 0.01         \\
              & \textbf{V} & 1778          & 201          & 1280          & 277          &  & 0.91         & 0.94          & 0.36            & 0.38         \\
              & \textbf{Q} & 451          & 77          & 25          & 2445          &  & 0.96         & 0.99          & 0.82         & 0.64        \\ \hline
              \midrule
\end{tabular}
\end{table}

\subsection{User Verification and Identification with ECG signals}

\begin{table}[!t]
\begin{center}
\caption{Performance on User Verification Task}
\label{tab:MITBIH_verif}
\begin{tabular}{cccccc}
\hline
\textbf{Model} &  & \multicolumn{4}{c}{\textbf{Performance}}                                                                           \\ \cline{3-6} 
                        &  & \textit{\textbf{Accuracy}} & \textit{\textbf{Specificity}} & \textit{\textbf{Sensitivity}} & \textit{\textbf{PPV}} \\ \hline
ResNet  & & 0.993 & 0.995 & 0.811 & 0.882     \\
\textbf{ECGBERT}  & & \textbf{0.999} & \textbf{1.000} & \textbf{1.000} & \textbf{0.972}     \\

\hline
\bottomrule
\end{tabular}

\end{center}
\end{table}

Besides the aforementioned cardiovascular-related tasks, we also compare a downstream task that is not related to cardiovascular disease, user authentication. Previous studies have demonstrated that ECG signals can serve as a biometric authentication method \citep{odinaka2012ecg,labati2019deep}. We analyze our results on the MIT-BIH arrhythmia dataset with 48 patient records. We label the data according to the user ID. We fine-tune the model for 20 epochs with a learning rate of 1e-4 and a batch size of 128. On a held-out test set, ECGBERT achieved an accuracy of 0.999 and a PPV of 0.972, as compared to an accuracy of 0.993 and a PPV of 0.882 for the ResNet, as it can be seen from Table \ref{tab:MITBIH_verif}.

Furthermore, we also test on a user identification task, where the model is to predict which user a certain ECG is from. Once again, ECGBERT outperforms the ResNet baseline as in Table \ref{tab:MITBIH_identif}. Due to the dataset being balanced across all classes, we only report the test accuracy for both models.

\begin{wraptable}{r}{5cm}
\begin{center}
\vspace{-25pt}
\caption{Performance on User Identification Task}
\vspace{6pt}
\label{tab:MITBIH_identif}
\begin{tabular}{cc}
\hline
\textbf{Model} &  \textit{\textbf{Accuracy}} \\ 
\hline
ResNet  & 0.920 \\
\textbf{ECGBERT}  & \textbf{0.938}    \\
\bottomrule
\hline
\end{tabular}
\end{center}
\vspace{-20pt}
\end{wraptable}

\subsection{Sleep Apnea Detection}
Obstructive sleep Apnea (OSA) plays a crucial role in health because it can potentially cause life-threatening problems such as heart failure or cognitive impairments \citep{beaudin2021cognitive}. Therefore, we chose to explore sleep Apnea detection with the PhysioNet Apnea-ECG Database v1.0.0 \citep{penzel2000Apnea}. We fine-tuned the ECGBERT with two extra dense layers with a batch size of 64 and a learning rate of 0.005, respectively for 5 epochs. From Table \ref{tab:Apnea_perf}, our baseline ResNet model achieved an accuracy of 0.709, a specificity of 0.744, a sensitivity of 0.653, and a PPV of 0.606. On the other hand, ECGBERT achieved an accuracy of 0.725, specificity of 0.626, sensitivity of 0.831, and PPV of 0.678. We believe the low performance is due to the noisiness of the signals compared with other datasets, making it difficult for the model to learn the specific representations. Moreover, since the length of the input data is 60 seconds, it makes it even more challenging since we only feed in 10-second sub-sequences of the Apnea data, while the labels are given per 60-second segment. Compared to the ResNet baseline, ECGBERT shows lower specificity but much higher sensitivity, along with higher overall accuracy.  Sensitivity, where correctly identifying positive instances, is a very critical concept in a medical study because high sensitivity demonstrates that the model is effective in capturing AFIB cases correctly.

\begin{table}[]
\centering
\caption{Inter-patient Performance on Sleep Apnea Detection}
\label{tab:Apnea_perf}
\begin{tabular}{ccccc}
\hline
\multirow{2}{*}{\textbf{Model}} & \multicolumn{4}{c}{\textbf{Performance}}                          \\ \cline{2-5} 
 & \textit{\textbf{Accuracy}} & \textit{\textbf{Specificity}} & \textit{\textbf{Sensitivity}} & \textit{\textbf{PPV}} \\ \hline
ResNet                          & 0.709          & \textbf{0.744} & 0.653          & 0.606          \\
\textbf{ECGBERT}                         & \textbf{0.725} & 0.626          & \textbf{0.831} & \textbf{0.678} \\ \hline
\bottomrule
\end{tabular}
\end{table}

\section{Limitations and Future Work}
We acknowledge that this work has some notable limitations. First, we do not conduct more experiments such as an ablation study, comparison of model parameters, or comparison with other works, which leaves the results with room for improvement.

Second, pre-training ECGBERT to learn more general representations of the ECG signals requires much more time and computing. Since this is a proof-of-concept study to whether the ECG signals can be interpreted as a language, we only trained on a minor subset, about 236 hours. In addition, we only employ the most widely-adopted lead-II ECG signals due to the fact that multi-lead downstream datasets currently are few and far between, and different conventions would require that ECGBERT fits all these paradigms. Further modifications could possibly be made to ECGBERT to take in a variable number of leads, with a specialized masking method so that the transformer only uses the non-masked signals. This could result in a multifaceted and more robust language model, though at the same time, it would require a vastly greater amount of computing. 

Finally, ECG language processing steps in Section \ref{sec:elp} require  the segmentation model to be able to correctly segment the ECG signal into appropriate small waves. In the presence of low-quality ECG signals, the wave segmentation module inconsistently and inaccurately segments the small waves for the abnormal ECG signals, which results in noisier outputs that affect the ECGBERT performance. This can be resolved with a more robust segmentation model with deep learning approaches to help better segment the noisy and abnormal ECG signals.

\section{Conclusion}
In this study, we propose the ECGBERT, a novel framework that can interpret the ECG signals and perform different downstream tasks with a single pre-trained model. The proposed approach consists of two main steps: 1) ECG language processing, and 2) large language modeling. After ECGBERT is pre-trained with a large amount of unlabeled data, it can be deployed to any downstream task that uses ECG signals as inputs. As we hypothesized, ECGBERT successfully learns general representations of different types of ECG signals, as shown by its stable performance on a variety of downstream tasks. We view this paper as the first foray into effectively utilizing a large amount of unlabeled, uncurated ECG data present online to solve various tasks efficiently without relying on internal and restricted datasets. This provides a non-block-box, repeatable model framework which allows for more accountability and repeatability on the part of deep learning researchers in the medical domain.


\newpage
\bibliography{sources}

\newpage
\appendix
\section{Appendix}
\subsection{Detailed Experimental Setup}
\subsubsection{Detailed Descriptions for Pretraining Datasets}
\begin{itemize}
    \item \textbf{MIMIC-III Matched Subset Dataset} includes 22,317 waveform records and 22,247 records for 10,282 distinct intensive care unit (ICU) patients. With a sample rate of 125 Hz, each record contains up to eight channels of signals such as ECG, arterial blood pressure (ABP), or PPG. It also contains additional contextual values such as respiration rate, SpO2, and blood pressure. We used the ECG lead-II signals for pre-training ECGBERT after resampling the signals to 250 Hz and splitting the signals into multiple 10-second segments. Given our limited computational resources, we randomly selected a subset of 13 records from the dataset. Correspondingly, within each selected record, we further employed a random sampling technique to select 3,000 segments to be used in the ECGBERT pre-training phase.

    \item \textbf{PTB-XL Dataset} contains 21,837 clinical 12 lead ECG signals from 18,869 patients with a length of 10 seconds. This dataset provides two different sampling rates at 100 Hz and 500 Hz and was annotated by two cardiologists, resulting in detailed diagnosis information across five different classes per record: Normal, Myocardial Infarction, ST/T Change, Conduction Disturbance, and Hypertrophy. During the pre-training phase, lead II ECG signals with a sampling rate of 500 Hz were chosen and then resampled to a rate of 250 Hz.

    \item \textbf{CPSC Dataset} has 6,877 12-lead ECG records with a 500 Hz sampling frequency. The duration of these records ranges from 6 to 60 seconds. The dataset contains nine different types of cardiac states, including atrial fibrillation (AFIB), ST-segment elevation (STE), ST-segment depression (STD), premature ventricular contraction (PVC), premature atrial contraction (PAC), normal heartbeat (Normal), left bundle branch block (LBBB), right bundle branch block (RBBB), and intrinsic paroxysmal atrioventricular block (IAVB). To facilitate the pre-training of the ECGBERT, any records exceeding a duration of 10 seconds were subdivided into multiple segments, each segment spanning 10 seconds.

    \item \textbf{Georgia Dataset} contains 20,672 ECG records where the duration of each record is between 5 and 10 seconds long with a 500 Hz sampling rate. After the records were resampled to 250 Hz, we only selected the 10-second ECG records for pre-training the ECGBERT.
    
\end{itemize}
\subsubsection{Detailed Descriptions for Downstream Task Datasets}
\begin{itemize}
    \item \textbf{MIT-BIH Atrial Fibrillation Dataset} contains 23 ECG recordings which were recorded for 10 hours and consist of two ECG signals, ECG1 and ECG2, sampled at 250 Hz with 12-bit resolution over a range of 10 millivolts, respectively. This dataset contains five rhythm classes that are manually annotated: atrial fibrillation (AFIB), atrial flutter (AFL), AV junctional rhythm (J), and normal (N) rhythms. We label AFIB as one and the remaining labels as zero to make this a binary classification task. We excluded 5 out of the 25 recordings due to missing signals. Herein, we used the ECG1, corresponding to lead-2, with a resampled frequency of 250 Hz and split each ECG signal into data segments of 10 seconds, and annotated a label for each of the data segments based on the majority voting system. If a data segment contained more AFIB heartbeats in 10-second ECG signals, it is labeled as AFIB, otherwise, it was labeled as non-AFIB arrhythmia. Please note that other downstream task datasets follow the same resample rate and labeling process otherwise mentioned explicitly.
    
    \item \textbf{MIT-BIH Arrhythmia Dataset} includes various beat types from 48 records of 47 subjects between 1975 and 1979 from the BIH Arrhythmia Laboratory. ECG recording duration is 30 minutes with 360 Hz sampling rate and a bandpass filter is applied to the signal at 0.1 Hz -- 100 Hz. In this study, we leveraged a modified limb lead-II signal, where each record has two channels (i.e., II or occasionally V5). Specifically, we split the data into multiple segments where each segment is 10 seconds of ECG signals and categorized different beat types into five classes, which are normal beat (N), ventricular ectopic beat (V), supraventricular ectopic beat (S), fusion beat (F), and unknown beat (Q) using a mixture of features.

    \item \textbf{Apnea-ECG Dataset} is comprised of 70 recordings, which are split into two sets: 35 records for the model training and another 35 for the testing. These recordings are available on the PhysioNet website. The ECG signals were captured at a sample rate of 100 Hz, offering a 16-bit resolution with 200 A/D units per mV. Each record spans a duration of approximately seven to ten hours, and annotations are provided to indicate normal and apnea occurrences per minute. In this study, the data was segmented into multiple non-overlapping segments, with each segment representing a duration of 10 seconds.

\end{itemize}

\label{sec:downstream_data}

- 

\end{document}